\documentclass[sigconf]{acmart}
\AtBeginDocument{%
  }

\usepackage{listings}
\usepackage{xcolor}

\usepackage{enumitem}
\setlist[itemize,1]{left=0pt}

\lstdefinelanguage{json}{
    basicstyle=\ttfamily\footnotesize,
    numbers=left,
    numberstyle=\tiny\color{gray},
    stepnumber=1,
    numbersep=5pt,
    showstringspaces=false,
    breaklines=true,
    frame=single,
    backgroundcolor=\color{gray!10},
    literate=
     *{0}{{{\color{blue}0}}}{1}
      {1}{{{\color{blue}1}}}{1}
      {2}{{{\color{blue}2}}}{1}
      {3}{{{\color{blue}3}}}{1}
      {4}{{{\color{blue}4}}}{1}
      {5}{{{\color{blue}5}}}{1}
      {6}{{{\color{blue}6}}}{1}
      {7}{{{\color{blue}7}}}{1}
      {8}{{{\color{blue}8}}}{1}
      {9}{{{\color{blue}9}}}{1}
      {:}{{{\color{black}:}}}{1}
      {,}{{{\color{black},}}}{1}
      {"}{{{\color{red}"}}}{1},
}

\lstdefinelanguage{yaml}{
  basicstyle=\ttfamily\footnotesize,
  numbers=left,
  numberstyle=\tiny\color{gray},
  stepnumber=1,
  numbersep=5pt,
  showstringspaces=false,
  breaklines=true,
  frame=single,
  backgroundcolor=\color{gray!10},
  morecomment=[l]{\#},
  commentstyle=\color{gray},
  morekeywords={true,false,null},
  keywordstyle=\color{blue},
  sensitive=false
}

\lstdefinelanguage{log}{
  basicstyle=\ttfamily\footnotesize,
  numbers=left,
  numberstyle=\tiny\color{gray},
  stepnumber=1,
  numbersep=5pt,
  showstringspaces=false,
  breaklines=true,
  frame=single,
  backgroundcolor=\color{gray!10},
  morecomment=[l]{//},
  commentstyle=\color{gray}
}

\lstdefinelanguage{dot}{
  morekeywords={digraph,strict},
  sensitive=false,
  morecomment=[l]{//},
  morestring=[b]",
  basicstyle=\ttfamily\footnotesize,
  numbers=left,
  numberstyle=\tiny\color{gray},
  stepnumber=1,
  numbersep=5pt,
  showstringspaces=false,
  breaklines=true,
  frame=single,
  backgroundcolor=\color{gray!10},
  keywordstyle=\color{blue},
  commentstyle=\color{gray}
}

\begin{document}

%%
%% The "title" command has an optional parameter,
%% allowing the author to define a "short title" to be used in page headers.
\title{Enabling Scientific Workflow Scheduling Research in Non-Uniform Memory Access Architectures}

%%
%% The "author" command and its associated commands are used to define
%% the authors and their affiliations.
%% Of note is the shared affiliation of the first two authors, and the
%% "authornote" and "authornotemark" commands
%% used to denote shared contribution to the research.
\author{Aurelio Vivas}
% \authornote{Both authors contributed equally to this research.}
\email{aa.vivas@uniandes.edu.co}
% \orcid{1234-5678-9012}
% \author{Harold Castro}
% \authornotemark[1]
% \email{hcastro@uniandes.edu.co}
\affiliation{%
  \institution{Universidad de los Andes}
  \city{Bogotá}
  % \state{Ohio}
  \country{Colombia}
}

\author{Harold Castro}
\email{hcastro@uniandes.edu.co}
\affiliation{%
  \institution{Univeersidad de los Andes}
  \city{Bogotá}
  \country{Colombia}}

%%
%% By default, the full list of authors will be used in the page
%% headers. Often, this list is too long, and will overlap
%% other information printed in the page headers. This command allows
%% the author to define a more concise list
%% of authors' names for this purpose.
\renewcommand{\shortauthors}{Vivas and Castro}

%%
%% The abstract is a short summary of the work to be presented in the
%% article.
\begin{abstract}
In recent years, data-intensive scientific workflows have increasingly adopted high-performance computing (HPC) systems as execution environments, complementing traditional Grid and Cloud platforms. However, HPC systems pose unique challenges for workflow scheduling due to their reliance on non-uniform memory access (NUMA) architectures.
% to scale computational performance. 
This architecture requires schedulers to consider not only data locality across the entire distributed infrastructure but also within individual computing nodes.
Modern HPC nodes feature complex memory hierarchies composed of multiple NUMA domains, each with heterogeneous memory regions such as high-bandwidth memory (HBM) and traditional DRAM. In many systems, specific NUMA nodes are also directly connected to accelerators, such as GPUs or FPGAs, with their own distinct memory spaces, as well as to network interface cards (NICs) with dedicated buffers. This architectural design complicates data access, as access latency varies significantly depending on the physical placement of both computation and data within the node.
Despite these challenges, most existing workflow scheduling algorithms were developed for Grid or Cloud environments and rarely address the complexities of NUMA-aware execution on HPC systems. To bridge this gap, this work introduces \textbf{nFlows}, a NUMA-aware Workflow Execution Runtime System that supports the modeling, bare metal execution, simulation, and validation of scheduling algorithms for data-intensive workflows on NUMA-based HPC systems.
We describe the system’s design and implementation, along with the functional tests used for validation. \textbf{nFlows} enables developers to build accurate simulation models, facilitating bare-metal execution of these models for comparison, studying NUMA effects in scheduling, designing NUMA-aware algorithms, analyzing data movement patterns, identifying performance bottlenecks, and studying in-memory workflow execution.
\newpage
\end{abstract}

%%
%% The code below is generated by the tool at http://dl.acm.org/ccs.cfm.
%% Please copy and paste the code instead of the example below.
%%
\begin{CCSXML}
<ccs2012>
   <concept>
       <concept_id>10011007.10011006.10011066.10011070</concept_id>
       <concept_desc>Software and its engineering~Application specific development environments</concept_desc>
       <concept_significance>500</concept_significance>
       </concept>
 </ccs2012>
\end{CCSXML}

\ccsdesc[500]{Software and its engineering~Application specific development environments}

%%
%% Keywords. The author(s) should pick words that accurately describe
%% the work being presented. Separate the keywords with commas.
\keywords{Runtime System, Tracing, Scientific Workflow, Scheduling
}
%% A "teaser" image appears between the author and affiliation
%% information and the body of the document, and typically spans the
%% page.
% \begin{teaserfigure}
%   \includegraphics[width=\textwidth]{sampleteaser}
%   \caption{Seattle Mariners at Spring Training, 2010.}
%   \Description{Enjoying the baseball game from the third-base
%   seats. Ichiro Suzuki preparing to bat.}
%   \label{fig:teaser}
% \end{teaserfigure}

% \received{20 February 2007}
% \received[revised]{12 March 2009}
% \received[accepted]{5 June 2009}

%%
%% This command processes the author and affiliation and title
%% information and builds the first part of the formatted document.
\maketitle

\section{Introduction}

In recent years, supercomputers have become essential for executing data-intensive scientific workflows. 
Despite this, the development of scheduling algorithms for these systems, as well as the tools that support their development, are still in their early stages.
As early as 2007, the rapid growth in scientific data, especially from supercomputer simulations, began to exceed the capacity for transferring data to remote platforms such as the Grid or the cloud for post-mortem analysis \cite{gray2007escience, da2024workflows}. This challenge led to strategic efforts by 2008 \cite{bergman2008exascale} and 2013 \cite{chen2013synergistic} to integrate high-performance computing (HPC) systems into the workflow life cycle, expanding their role from data generation to in-situ data processing \cite{yildiz2017efficient, deelman2018future}. 
The launch of Frontier in 2022, the first exascale system with Big Data processing capabilities, underscored this expanded role \cite{dongarra2022report}.
While several scientific workflows are now executed on HPC systems, they typically rely on traditional scheduling algorithms originally developed for cloud or grid environments \cite{vivas2024trends, cid2020efficient}; moreover, these algorithms often overlook critical architectural features, such as the non-uniform memory access (NUMA) design common in modern supercomputers, leading to suboptimal performance \cite{amela2017enabling}.

Non-uniform memory access (NUMA) architectures present distinct challenges for executing scientific workflows.
First, data access times and memory bandwidth vary across cores. Cores with local access to a memory region experience lower latencies and higher bandwidth, while accessing remote memory significantly increases latency and reduces bandwidth.
Second, task placement becomes critical. Tasks that frequently communicate or synchronize benefit from minimized physical distance between them. Conversely, independent, memory-intensive tasks may perform better when assigned to different memory partitions, thus distributing memory traffic across the NUMA system \cite{broquedis2010hwloc}.
Finally, the operating system can significantly influence execution performance through mechanisms such as core migrations, NUMA balancing, data page distribution, hardware and compiler prefetching, and thread memory allocation policies.

Non-uniform Memory Access (NUMA) architectures have been thoroughly investigated in the context of numerical applications. 
Researchers have meticulously analyzed these applications, including those based on dense and sparse linear algebra, n-body methods, structured and unstructured grids, and Monte Carlo simulations, to pinpoint their computational and data movement patterns like one-to-one, one-to-all, all-to-all, gather, and scatter \cite{asanovic2006landscape, asanovic2009view}. This deep understanding has even led to practical tools such as Portable Hardware Locality (hwloc), empowering developers to map tasks and memory regions for optimal locality exploitation \cite{broquedis2010hwloc, goglin2014managing, goglin2017overhead}.
However, the impact of NUMA on scientific workflow applications remains largely unexplored. 
While these applications have focused on identifying benchmark workflows, such as Montage, Cybershake, LIGO, and Epigenomics \cite{juve2013characterizing}, and describing their high-level data processing patterns (e.g., pipelines, data distribution, data aggregation, and data redistribution) \cite{bharathi2008characterization}, there's a significant lack of research into their performance on NUMA-based HPC systems and how memory access patterns affect their execution.
This analytical gap severely hinders our ability to design efficient schedulers that effectively account for NUMA characteristics in scientific workflows.

This work presents \textbf{nFlows} (\textbf{N}UMA-Aware Work\textbf{flow} Execution Runtime \textbf{S}ystem) \cite{aurelio_vivas_2025_15811369}, a C/C++ program built to model, execute (both in simulation and on actual hardware), and analyze scientific workflow scheduling algorithms, specifically considering NUMA (Non-uniform Memory Access) effects in HPC systems.
\textbf{nFlows} \cite{aurelio_vivas_2025_15811369} emulates the execution of workflows with clearly defined structures. It models computational tasks using floating-point operations (FLOPs) and communication payloads in bytes. During execution, it traces tasks and data locality, and records data access performance. This collected data can be accessed by dynamic scheduling algorithms at runtime to improve scheduling decisions.
The system leverages POSIX threads (pthreads) for asynchronous task execution and core binding, and the Portable Hardware Locality (hwloc) library to monitor task placement and data locality across NUMA nodes.

\section{Related Work}

The impact of non-uniform memory access (NUMA) architectures has been extensively studied in numerical parallel applications using simulators and runtime systems. In contrast, to the best of our knowledge, no prior work has proposed a runtime system that enables modeling, simulation, bare-metal execution, and analysis of scientific workflow scheduling under NUMA effects in high-performance computing (HPC) systems.
This gap can be attributed to two main reasons. First, numerical applications have traditionally served as the primary use cases for HPC systems, often taking precedence over data-intensive scientific workflows. Second, although the introduction of tasking models in parallel programming frameworks has enabled the representation of parallel applications as directed acyclic graphs (DAGs), these models typically focus on structured parallel patterns, such as stencil computations, which differ significantly from scientific workflows in terms of task granularity and graph structure.
The following sections describe the most closely related work.

% Runtime systems
% Machado et al. (2023) \cite{machado2023source},
% Dokulil and Benker (2020) \cite{dokulil2020automatic} numerical parallel applications, in particular stencil patterns
% Jamil et al. (2023) \cite{jamil2023throughput},  optimizing I/O thorughout in stream based scientific workflows specific use case. their work emphasizes optimizing CPU core usage and maximizing network I/O throughput.

% Simulators
% Daoudi et al. (2020) \cite{daoudi2020somp} OpenMP, numerical parallel applications.
% Liu et al. (2013) \cite{liu2013simnuma}
% Tao et al (2004) \cite{tao2003simulation, tao2004simt} numerical parallel applications.

\subsection{Runtime Systems}

Machado et al. (2023) \cite{machado2023source} proposed a source-to-source compiler to instrument C code for tracking the type of each memory access (local or remote) performed by a numerical parallel program executed in a NUMA-based system, a profiler to collect memory access information and provide detailed statistics, and a metric to evaluate the quality of memory access patterns.
For each memory access, the profiler records the virtual memory address, the NUMA node ID of its physical location, and the NUMA node ID of the thread making the request. To gather NUMA-related information and control the execution of the program's threads, the authors utilized \texttt{libnuma} to identify the NUMA node ID of a given memory address and the NUMA node ID where the thread is running. Additionally, they used \texttt{numactl} to bind thread execution and memory allocation to specific NUMA nodes, ensuring that the locality policy (thread and data) is set at the command level and inherited by all child processes.
% \textcolor{blue}{
% Although Machado et al.'s work shares similarities with ours, such as the collection of locality data, there are two significant differences between their methodology and ours. First, their approach targets numerical parallel applications, particularly stencil computations like wave simulation codes. Second, their methodology requires the definition of lexical patterns to generate the source-to-source compiler capable of instrumenting the identified patterns in the code.
% }

Jamil et al. (2023) \cite{jamil2023throughput} proposed a runtime system to optimize the utilization of available CPU cores and minimize network I/O during the execution of data stream-based scientific workflows. Their work focuses on workflows like those integrating the Advanced Photon Source (APS) at Argonne National Laboratory (ANL) with High-performance computing (HPC) systems at the Argonne Leadership Computing Facility (ALCF).
The proposed system executes the workflow across three computing infrastructures: the APS, where scientific data is generated; an upstream gateway node responsible for pre-processing tasks such as aggregation and load balancing; and the HPC system, where the processed data is further analyzed. The authors concentrated on the streaming process between the APS and the upstream gateway node by mimicking the data generation machine (APS) and the data receiving machine (gateway node) in a prototype environment using NUMA-based machines.
To enhance network I/O throughput between these two machines, the runtime system leverages the \textit{libnuma} API to set NUMA affinity for workflow tasks, overriding the operating system’s default scheduling decisions, which are not NUMA-aware. Compression tasks on the data generation machine were assigned to underutilized cores, increasing compression speed and improving data emission rates. On the receiving machine, decompression threads were allocated to the NUMA node directly connected to the network interface card (NIC), resulting in a 15\% improvement in overall throughput. 
% \textcolor{blue}{
% While the authors utilized the \texttt{libnuma} library to bind threads and achieve NUMA affinity and analyzed the average memory access for each core, their work differs significantly from ours in two key aspects. First, their focus is on data stream-based scientific workflows rather than batch-based workflows. Second, their work emphasizes optimizing CPU core usage and maximizing network I/O throughput when leveraging NUMA memory architectures, adopting an infrastructure-centered approach. In contrast, our research aims to model and analyze the overall performance of workflows, adopting an application-centered approach.
% }

Dokulil and Benker (2020) \cite{dokulil2020automatic} proposed a runtime system comprising three components: a profiler that collects memory access information from numerical parallel applications, an analyzer that performs post-mortem analysis on the collected data to generate an improved execution plan, and modifications to the OCR-Vx NUMA-aware scheduler to load the plan and execute the application with enhanced performance.
The runtime system optimizes work and data placement under the assumption that it will be applied to applications with a consistent structure, such as iterative stencil computations. For example, when the runtime profiles an iterative stencil application during its initial execution, it generates patterns that can be reused in subsequent runs of the same application, even with different input parameters, as long as the application's execution structure remains unchanged.
% \textcolor{blue}{
% While the authors provide a scheduling mechanism for executing workflows (in this case, task dependency graphs (TDG) representing numerical parallel applications), collect memory access information during execution, and analyze this information further, their work has significant differences compared to ours. First, their approach is limited to numerical parallel applications. Although the execution abstractions (e.g., task dependency graphs) resemble workflows, they differ substantially from scientific workflows. Second, the runtime system is tailored specifically for iterative stencil computations, which may render structured task dependency graphs. In contrast, scientific workflows involve more complex task execution and data movement patterns that cannot be represented in stencil-like parallel applications and, therefore, be analyzed with this profiler. Finally, the authors do not specify the details of the memory access data collected during the execution of parallel threads, limiting the understanding of how memory behavior is characterized.
% }

\subsection{Simulators}

Daoudi et al. \cite{daoudi2020somp} (2020) developed sOMP, a SimGrid-based simulator designed to model the effects of contention and data access in NUMA architectures. The primary goal of sOMP is to predict the performance of task-based parallel applications under these conditions. The model includes processing units, memory, and their interconnections but excludes the effects of L1/L2 caches and does not exhaustively model memory topology.
Processing units (cores) were mapped onto SimGrid hosts, with the memory controller (also a host) acting as an intermediary for communications without performing computations. Interconnections were modeled using SimGrid routers and links. Hosts (cores) were grouped into NUMA nodes and sockets according to the real architecture, with links connecting these groups. Parameters such as latency and bandwidth were set based on real machine characteristics, measured using BenchIT, x86membench, and Intel Memory Latency Checker.
The model's accuracy was evaluated by comparing predicted execution times with those obtained from real Intel and AMD machines. Task graphs for parallel applications were constructed from traces captured from dense linear algebra applications with varying data sizes, using the proposed TiKKi tracing tool built on the OMPT API. Two task graph models were evaluated: one without data transfer costs and another that included data transfer costs. The simulator demonstrated small relative errors across Intel and AMD architectures and achieved faster simulation times compared to real-system execution.
This work was later extended in \cite{daoudi2023improving} to consider NUMA and cache effects.

Liu et al. (2013) \cite{liu2013simnuma} introduced SimNUMA, an execution-driven simulator designed to model entire clusters, including their NUMA memory architectures. The primary objective of SimNUMA is to estimate the execution time of applications on target machines accurately.
SimNUMA calculates execution time as the sum of instruction execution time ($T_{inst}$), remote memory access time ($T_{rmem}$), I/O wait time ($T_{io}$), and ready wait time ($T_{ready}$). By intercepting program instructions, remote memory accesses, process scheduling, inter-process communication, and file access operations, SimNUMA estimates these time components.
The simulator's accuracy was evaluated by comparing its predicted execution times with those obtained on real target machines using benchmark applications from SPEC CPU2006 and SPLASH-2. The largest system modeled consisted of 64 computing nodes. At the beginning of its operation, the configuration module reads parameters from a configuration file, specifying details like the number of processors, memory size, and interconnection network characteristics.
% \textcolor{blue}{No designed for the development and assessment of scheduling algorithms, nor scientific workflow applications.}

Tao et al. (2004) \cite{tao2004simt, tao2003simulation} introduced SIMT, a simulator designed to evaluate the performance of memory systems, particularly focusing on cache behavior (cache misses, cache line invalidation) and memory locality in NUMA architectures. SIMT primarily models caches, distributed shared memory, and data transfer between processor nodes.
SIMT simulates the parallel execution of shared memory applications on multiprocessor systems.
It is composed of a front-end that simulates the activity of multiple processes across processors and captures relevant events, while the back-end processes these events and models the actions that would occur on the target system.
The simulator incorporates multi-level cache modeling and supports various cache coherence protocols, such as MESI and Optimal, along with data allocation strategies, including round-robin.
Communication latencies in the system are modeled using fixed values.
SIMT collects a comprehensive range of statistics during simulation, including total simulation time, the number of processor cycles simulated, memory references made, and detailed data on cache hits and misses for each level of cache in the architecture.
The simulator has been applied to test workloads from numerical applications such as FFT, LU, RADIX, and WATER.
To validate SIMT, the authors conducted a detailed comparison between SIMT's outputs and real hardware performance using multiple performance metrics. The study compared both absolute performance values and relative performance trends.
Results demonstrated that SIMT was generally effective at providing accurate performance predictions, particularly for identifying performance trends.

\section{NUMA-Aware Workflow Execution Runtime}

\begin{figure*}[ht]
    \centering
    \includegraphics[width=\linewidth]{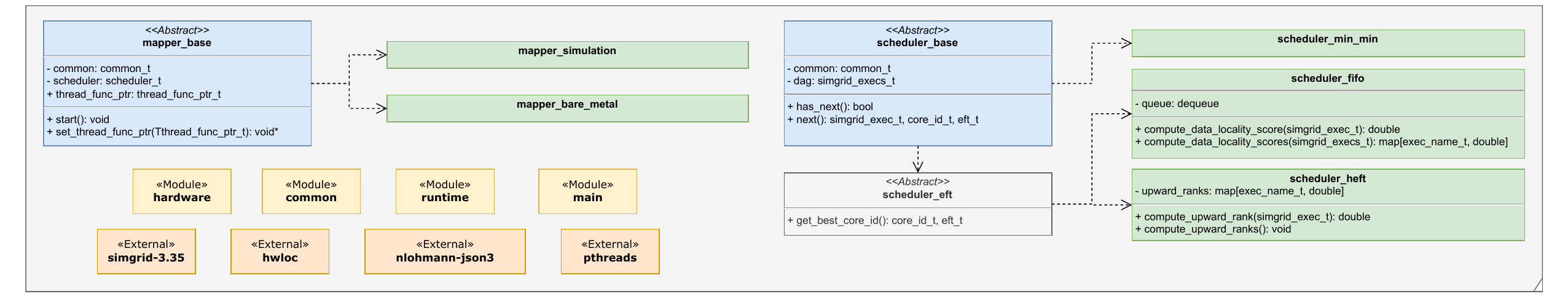}
    \caption{nFlows' Class Diagram}
    \label{fig:nflow_class_diagram_h}
\end{figure*}

Figure~\ref{fig:nflow_class_diagram_h} presents the \textbf{nFlows} \cite{aurelio_vivas_2025_15811369} class diagram, which illustrates the main components and external dependencies of the system. It includes four modules (hardware, common, runtime, and main), three abstract classes (mapper\_base, scheduler\_base, and scheduler\_eft), and five concrete implementations (mapper\_simulation, mapper\_bare\_metal, scheduler\_fifo, scheduler\_heft, and scheduler-min\_min).

The \textbf{hardware module} defines interfaces to query core and memory locality using \texttt{hwloc} and to bind threads to specific cores via the \texttt{pthreads} library. 
The \textbf{common module} provides the data structure that maintains the runtime system state, including DAG execution state, distance matrices (latency and bandwidth), core availability, and tracing data. This data structure is only accessible through thread-safe interfaces. 
The \textbf{runtime module} encapsulates the control logic of the system and exposes four functions: initialize, finalize, start, and stop. The \textbf{main module} serves as the runtime entry point.

\textbf{Mappers} are responsible for assigning workflow tasks to the target infrastructure (simulated or real). They query the scheduler for the next task to execute and handle its execution, including reading/writing data and performing computations. The simulation mapper uses basic models to estimate execution and communication time, while the bare-metal mapper executes these operations on the real system using predefined routines that mimic actual resource usage.

% \textbf{Schedulers} implement the logic of the scheduling algorithms, in this case, NUMA-aware versions of Min-Min, FIFO, and HEFT. Each scheduler must define two functions: has\_next(), which indicates whether there are remaining tasks to schedule, and next(), which returns a tuple with the next task, the selected core ID, and the estimated earliest completion time. A null task signals that no ready tasks are available, while a null core ID indicates that no cores are currently free. 
% Both signals indicating the mapper has to wait until new tasks and cores are available.
\textbf{Schedulers} implement the logic of the scheduling algorithms, in this case, NUMA-aware variants of Min-Min, FIFO, and HEFT. 
Each scheduler must define two functions: \texttt{has\_next()}, which indicates whether there are remaining tasks to schedule, and \texttt{next()}, which returns a tuple containing the next task, the selected core ID, and the estimated earliest completion time. A \texttt{null} task indicates that no ready tasks are currently available, while a \texttt{null} core ID signifies that no cores are free at the moment.
Both conditions signal to the mapper that it must wait until new tasks become ready and cores become available.
Further details on the implemented algorithms and the runtime validation are provided in Section \ref{sec:validation}.

\subsection{Workflow Model}

The workflow is modeled as a directed acyclic graph (DAG) \( G = (V, E) \), where \( V = \{v_1, v_2, \dots, v_n\} \) is the set of tasks and \( E \subseteq V \times V \) represents data dependencies. Each task \( v_i \in V \) has a fixed computational cost defined by the number of floating-point operations required. Each edge \( (v_i, v_j) \in E \) denotes a data item produced by \( v_i \) and consumed by \( v_j \), with size \( d_{i,j} \) bytes. Thus, each task produces a distinct data item for each successor.
At task start, all input data \( (v_k, v_i) \in E \) must reside in memory, where they are consumed and then deallocated. The task then produces output data \( (v_i, v_j) \in E \), which are stored in memory for successors. This strict data handling ensures no task simultaneously holds both input and output data in memory, consistent with the data flow model in \cite{marchal2018parallel}.
\newline

% \noindent \textcolor{blue}{ Provide a graph comparing a sample workflow and its equivalent in the data flow model. Provide a data item (square) per dependency and provide information about the write and read costs as separate costs.}

\subsection{System Model}

The target computing system follows a non-uniform memory access (NUMA) architecture, consisting of $P$ homogeneous physical cores, organized into $M$ NUMA nodes. Each node $N_m$ (for $m = 1, \dots, M$) contains a set of cores and a dedicated local memory region $M_m$ with high-bandwidth, low-latency access for its resident cores. The memory model assumes direct main memory access, ignoring intermediate caches.
Each core $c_p \in N_m$ accesses data items from its
local memory with latency $L_m$ and bandwidth $B_m$. If a core $c_p \in N_m$ needs to access a data item $d_{i,j}$ located in a remote node $N_n$ (where $m \neq n$), the access incurs a higher latency  $L_{m,n}$ and is constrained by the interconnect bandwidth $B_{m,n}$ linking the nodes.
\newline

% \noindent \textcolor{blue}{ Provide figures describing the data model in NUMA systems. Describe data is partitioned into several pages and that it can be migrated between NUMA nodes. Describes how data is allocated in the system according to memory allocation policies. Emphasize in the memory models considered in the current implementation.}

\subsection{Input}
\label{sec:input}

\begin{figure}
\centering
\begin{lstlisting}[language=dot, caption={DAG specification in DOT format}, label={lst:dot_sample}]
strict digraph {
    v1 [size=10000000];
    v2 [size=10000000];
    v3 [size=10000000];
    v4 [size=10000000];
    v5 [size=10000000];
    v6 [size=10000000];

    root [size=10];
    end [size=10];

    root -> v1 [size=10];

    v1 -> v2 [size=60000000];
    v1 -> v3 [size=50000000];
    v1 -> v4 [size=40000000];

    v2 -> v5 [size=30000000];
    v3 -> v5 [size=20000000];

    v4 -> v6 [size=10000000];

    v5 -> end [size=10];
    v6 -> end [size=10];
}
\end{lstlisting}
\end{figure}

\begin{figure}
\centering
\begin{lstlisting}[language=json, caption={Configuration file for workflow execution}, label={lst:json_sample}]
{
    "dag_file": "./example/sample.dot",

    "scheduler_type": "fifo",
    "scheduler_params": [
        "fifo_prioritize_by_core_id=yes",
        "fifo_prioritize_by_exec_order=yes"
    ],

    "mapper_type": "bare-metal",
    "mapper_mem_policy_type": "first-touch",
    "mapper_mem_bind_numa_node_ids": [],

    "core_avail_mask": "0x100100",
    "flops_per_cycle": 32,
    "clock_frequency_type": "static",
    "clock_frequency_hz": 1000000000,

    "distance_matrices": {
        "latency_ns": "./example/non_uniform_lat.txt",
        "bandwidth_gbps": "./example/non_uniform_bw.txt"
    },

    "out_file_name": "./example/sample.yaml"
}
\end{lstlisting}
\end{figure}

Listings~\ref{lst:dot_sample} and~\ref{lst:json_sample} present a workflow specification in DOT format and its corresponding execution configuration.
In the DOT specification, the size attribute of each vertex denotes the computational payload in floating-point operations (FLOPs), whereas for edges, it indicates the communication payload in bytes. The workflow must define a unique entry (root) and a unique exit (end) node.
This requirement is imposed by the SimGrid library to facilitate workflow manipulation.
The execution configuration file specifies the input workflow, the scheduling algorithm to be applied (e.g., fifo, heft, min-min), algorithm-specific params, the mapping strategy (bare-metal or simulation), and various hardware-related parameters. These include the number of available cores, core performance and frequency, and NUMA distance matrices.
It also defines the name of the output file to be generated.

NUMA systems exhibit a variety of memory allocation policies that govern how data pages are distributed across memory regions, which are exposed to the system as a unified virtual address space~\cite{gaud2015challenges}.
Through mapper options, the runtime system allows specifying memory allocation policies (e.g., first-touch, next-touch, interleave) applied to asynchronous threads responsible for task execution.
It also constrains the memory regions accessible to these threads for data page allocation by specifying the hwloc logical IDs of the corresponding NUMA nodes.
It is important to note that, in simulation mode, the first-touch policy is assumed by default. Support for additional policies and their semantics within the simulation context must be explicitly defined.

Regarding computing resources, a bitmask is used to indicate which physical cores are enabled for task execution. The system can optionally incorporate flops per cycle and clock frequency information into scheduling decisions. The clock frequency may be specified as a single static value, as an array with one value per core, or dynamically retrieved at runtime.

\subsection{Output}
\label{sec:output}

\begin{figure}
\centering
\begin{lstlisting}[language=yaml, caption={Execution trace and metadata}, label={lst:yaml_sample}]
user:
  flops_per_cycle: 32
  clock_frequency_type: static
  clock_frequency_hz: 1e+09
  distance_lat_ns:
    - [67.9, 136.4]
    - [137.4, 68.8]
  distance_bw_gbps:
    - [120875, 34472]
    - [34471.5, 120849]

workflow:
  execs_count: 6
  reads_count: 6
  writes_count: 6

runtime:
  threads_checksum: 0
  threads_active: 0
  tasks_active_count: 6
  reads_active_count: 6
  writes_active_count: 6
  core_availability:
    8: {avail_until: 304488}
    20: {avail_until: 240743}

trace:
  name_to_thread_locality:
    v6: {numa_id: 0, core_id: 8, voluntary_cs: 0, involuntary_cs: 1, core_migrations: 0}
    v5: {numa_id: 1, core_id: 20, voluntary_cs: 0, involuntary_cs: 1, core_migrations: 0}
    ...
  numa_mappings_write:
    v4->v6: {numa_ids: [0]}
    ...
  numa_mappings_read:
    v4->v6: {numa_ids: [0]}
    ...
  comm_name_read_offsets:
    v4->v6: {start: 255820, end: 275216, payload: 1e+07}
    ...
  comm_name_write_offsets:
    v4->v6: {start: 249287, end: 255820, payload: 1e+07}
    ...
  exec_name_compute_offsets:
    v6: {start: 275216, end: 304488, payload: 1e+07}
    ...
    v1: {start: 0, end: 45259, payload: 1e+07}
    ...
  exec_name_total_offsets:
    v6: {start: 255820, end: 304488, payload: 1e+07}
    ...
    v1: {start: 0, end: 85235, payload: 1e+07}
\end{lstlisting}
\end{figure}

Listing~\ref{lst:yaml_sample} shows the output generated for the workflow defined in Listing~\ref{lst:dot_sample}, executed under the experimental configuration described in Listing~\ref{lst:json_sample}.
The runtime system maintains detailed data for validation purposes, as well as a trace of workflow execution activities, both computation and communication, along with their corresponding locality within the system.

For validation purposes, the runtime system reports:
\textit{user input data}, describing the parameters that may influence the behavior of the scheduling algorithm.
\textit{workflow-related information}, including the number of tasks and the number of read/write operations that must be performed. The total number of read and write operations must match to ensure that every data item allocated in memory is eventually deallocated after being consumed.
\textit{runtime validation data}, such as a checksum: the sum of all values read across memory read operations. Since the buffers are initialized to zero on every write operation, the sum of all read values must also be zero. Additionally, the runtime verifies that all threads spawned to execute individual tasks are correctly resumed upon completion. Counters are used to track the total number of tasks executed, and the number of read and write operations performed once the workflow has been fully scheduled. These values are updated concurrently by threads during execution, enabling the system to detect incomplete executions or unexpected terminations due to blocked or interrupted threads.

For scheduling algorithm evaluation, the trace includes detailed records of workflow execution:
\textbf{\textit{name\_to\_thread\_locality}},
specifies the NUMA node and CPU core where each task was executed. It also logs the number of voluntary and involuntary context switches, as well as core migrations. Core migrations should always be zero to ensure exclusive, non-preemptive access to CPU cores.
\textbf{\textit{numa\_mappings\_write/read}} records the locality, i.e., the NUMA node ID, of the memory regions holding each data item (e.g., $v_4 \rightarrow v_6$) involved in a communication.
Each data item transferred between tasks consists of multiple memory pages, whose size is defined by the operating system, and which may be allocated across one or more memory regions. When pages are distributed among multiple NUMA nodes, all corresponding node IDs are recorded; however, the trace does not indicate how many pages reside in each region.
Moreover, if a core initiates a read operation from a memory region located in a different NUMA node, the operating system may trigger page migration. As a result, even data initially allocated within a single NUMA node may be partially migrated during execution, modifying its effective locality.
This is why the runtime records the locality of every data item both after writing and after reading.
\textbf{\textit{comm\_name\_write/read\_offsets}}
records the start and end offsets, in microseconds ($\mu$s), of memory write and read operations, along with the size of the communicated payload in bytes.
\textbf{\textit{exec\_name\_compute\_offsets}}
records the start and end offsets, in microseconds ($\mu$s), of CPU-intensive operations, in this case, the execution of the fused multiply-add (FMA) kernel for the specified number of FLOPs.
\textbf{\textit{exec\_name\_total\_offsets}} records the actual start and end time offsets of a task. A task is considered to start when the first input data item is read, and to end when the last output data item is written to memory.

\subsection{Logging}

\begin{figure*}[ht]
\centering
\begin{lstlisting}[language=log, caption={Runtime log excerpt showing task scheduling and memory actions}, label={lst:log_sample}]
[runtime/INFO] Initialize
[runtime/INFO] Start
[mapper_bare_metal/INFO] Start
[fifo_scheduler/DEBUG] fifo_prioritize_by_exec_order: yes, enabled: true
[fifo_scheduler/DEBUG] priority_queued_task: v1, score: 0.000000
[fifo_scheduler/DEBUG] fifo_prioritize_by_core_id: yes, enabled: true
[fifo_scheduler/DEBUG] avail_core_id: 8, avail_core_until: 0.000000
[fifo_scheduler/DEBUG] avail_core_id: 20, avail_core_until: 0.000000
[fifo_scheduler/DEBUG] best_core_id: 8, best_numa_id: 0
[fifo_scheduler/DEBUG] selected_task: v1, selected_core_id: 8, estimated_finish_time: 313.064279
...
[mapper_bare_metal/INFO] Thread ID: 15445, Task ID: v1, Core ID: 8 => message: started.
[mapper_bare_metal/INFO] ... => thread_mem_policy: first-touch [0,1].
[mapper_bare_metal/INFO] ... => write: v1->v2, payload: 60MB, locality: [0].
[mapper_bare_metal/INFO] ... => write: v1->v3, payload: 50MB, locality: [0].
[mapper_bare_metal/INFO] ... => write: v1->v4, payload: 40MB, locality: [0].
[mapper_bare_metal/INFO] Thread ID: 15445, Task ID: v1, Core ID: 8 => message: finished.
...
[mapper_bare_metal/INFO] Thread ID: 15471, Task ID: v2, Core ID: 8 => message: started.
[mapper_bare_metal/INFO] ... => thread_mem_policy: first-touch [0,1].
[mapper_bare_metal/INFO] ... => read: v1->v2, payload: 60MB, checksum: 0
[mapper_bare_metal/INFO] ... => read: v1->v2, locality_before: [0], locality_after: [0], pages_migration: no
...
[mapper_bare_metal/INFO] Thread ID: 15471, Task ID: v2, Core ID: 8 => message: finished.
...
[0.060000] [mapper_bare_metal/INFO] End
[0.060000] [runtime/INFO] End
[0.060000] [runtime/INFO] Finalize
\end{lstlisting}
\end{figure*}

Listing~\ref{lst:log_sample} presents the execution logs for the workflow defined in Listing~\ref{lst:dot_sample}, using the configuration specified in Listing~\ref{lst:json_sample}.
These logs capture details related to task execution (thread identifiers and assigned core IDs), memory access events (reads, writes, and page migrations), NUMA locality, and thread-level activity.
In addition, they describe the interaction among the runtime system modules, including the runtime, scheduler, and mapper, throughout the execution of the workflow.

In the \textit{scheduler section}, the logs include information such as user-defined parameters passed through the configuration file and their validation, details of the scheduling algorithm (e.g., ready tasks in the priority queue and their associated scores), available cores and their expected availability time, the selected core and its NUMA locality, and the estimated finish time (EFT) of the scheduled task.

In the \textit{threads section}, the logs report the thread identifier, the core assigned to that thread, and the task mapped for execution. Additionally, the memory allocation policy for each thread is shown. For instance, thread 15445 is permitted to allocate memory pages in NUMA nodes 0 and 1. As detailed in Section~\ref{sec:input}, both thread-specific memory policies and the set of allowed memory regions for page allocation can be specified through the experiment configuration file.
The first task in the workflow (Listing~\ref{lst:dot_sample}), $v_1$, allocates three data items ($v_1 \rightarrow v_2$, $v_1 \rightarrow v_3$, and $v_1 \rightarrow v_4$) in the memory region associated with NUMA node 0. The subsequent task, $v_2$, reads its corresponding data item ($v_1 \rightarrow v_2$) and deallocates it.
Whenever a core requests a data page located in a remote memory region (i.e., one with a different locality), the operating system may trigger a page migration from the remote region to one that shares locality with the requesting core. Therefore, for each read operation, the runtime tracks the locality of the accessed data item both before and after the operation. If these localities differ, the event is classified as a page migration.

\section{Validation}
\label{sec:validation}

\textbf{nFlows} \cite{aurelio_vivas_2025_15811369} was validated following the credibility framework in \cite{blattnig2008towards}, focusing on code and solution verification. To assess runtime correctness, three traditional scheduling algorithms were extended with NUMA awareness, and functional tests were conducted to ensure that the system produces the expected outputs for each algorithm under specific configurations.

Test cases were developed for both the simulation mapper and the bare metal mapper:
the simulation mapper enables validation of the timing calculations produced by the simplified CPU and raw communication models, as well as the trace data collected during execution (i.e., computation and communication time offsets). Since the implemented scheduling algorithms produce deterministic results, such as workflow makespan, these outcomes can be accurately predicted and compared against expected values for basic workflow instances.
For the bare metal mapper, validation focuses on aspects such as selected CPU cores and task execution order, since execution timings depend on actual hardware performance and may vary between runs.

The trace output file (see Section~\ref{sec:output}) is validated using two Python scripts: \texttt{validate\_offsets.py} and \texttt{validate\_output.py}.
Because the trace format allows reconstruction of the workflow structure, \textit{validate offsets} checks that no task starts execution before all its dependencies are satisfied.
In turn, \textit{validate output} verifies that the generated trace matches the expected output for a given workflow and system configuration.

The following sections describe the implemented scheduling algorithms, validation scripts, and functional test cases in detail.

\subsection{Algorithms}
\label{sec:algorithms}

\subsubsection{Min-Min} 

The Min Min scheduling algorithm is a widely used heuristic for minimizing makespan in scientific workflow scheduling, especially common in cloud computing environments \cite{chen2013user, gupta2022enhanced}. It operates through three main phases: task initialization, task selection, and resource allocation.
During initialization, the algorithm computes the earliest finish time (EFT) of every unscheduled task on all available processing elements (e.g., CPU cores). In the selection phase, the task-resource pair yielding the minimum EFT is chosen, prioritizing the task that completes earliest overall. The allocation phase assigns the selected task to the chosen resource and updates its availability. This cycle repeats until all tasks are scheduled, with EFTs recalculated after each assignment.

Our implementation builds on the approach in \cite{Simulati23:online}.
This implementation accounts for heterogeneous bandwidths between communicating cores.
Communication cost is modeled as shown in Equation \ref{eq:comm_cost_original}.
Here, $L_{m}$ denotes the communication startup latency for the source processor $m$, and $B_{m,n}$ represents the available bandwidth between processors $m$ and $n$. 
This model assumes a direct processor-to-processor data exchange.
We extended this model in three ways: (1) incorporating NUMA concepts; (2) explicitly modeling non-uniform latencies; and (3) considering processor-memory interactions, as detailed in Equations \ref{eq:comm_cost_numa}, \ref{eq:est}, and \ref{eq:eft}.
Here, \( m \) and \( n \) denote NUMA domains, representing the locality of either computing cores or memory regions, rather than individual cores. 
The matrices \( L_{m,n} \) and \( B_{m,n} \) are symmetric and represent the latencies (in ns) and bandwidths (in GB/s), respectively, between NUMA domains. 
Finally, the EFT depends on the EST, the read time of the most time-consuming predecessor, the compute time, and the write time of the most time-consuming successor.

\begin{figure}[h]
\begin{equation}
c_{i,k} = L_m + \frac{\text{data}_{i,k}}{B_{m,n}}
\label{eq:comm_cost_original}
\end{equation}

\begin{equation}
c'_{i,k} = L_{m,n} + \frac{\text{data}_{i,k}}{B_{m,n}}
\label{eq:comm_cost_numa}
\end{equation}

\begin{equation}
EST'(t_i, p_j) = \max\left( avail[j], \max_{t_m \in pred(t_i)} AFT(t_m) \right)
\label{eq:est}
\end{equation}

\begin{equation}
EFT'(t_i, p_j) = EST'(t_i, p_j) + \max_{t_k \in pred(t_i)} { c'_{k,i} } + w_{i,j} + \max_{t_k \in succ(t_i)} { c'_{i,k} }
\label{eq:eft}
\end{equation}
\end{figure}

\subsubsection{Heterogeneous Earliest Finish Time (HEFT)}

HEFT \cite{topcuoglu2002performance} is a heuristic for mapping tasks in a directed acyclic graph (DAG) onto heterogeneous processors, aiming to minimize the overall completion time (makespan).
HEFT operates in three phases: initialization, task prioritization, and scheduling.
In the initialization phase, each task is assigned an average execution cost, computed as the mean of its execution times across all available processors. 
Similarly, dependencies between tasks are labeled with average communication costs, determined by averaging data transfer times between processor pairs. 
In the prioritization phase, tasks are ranked according to their upward rank, defined as the length of the longest path from the task to the exit node, incorporating both computation and communication costs.
Tasks are then sorted in descending order of this rank to determine their scheduling priority. 
In the scheduling phase, tasks are assigned one by one to the processor that yields the earliest finish time (EFT), considering the processor’s current availability and the arrival time of input data from predecessor tasks. 
The EFT is computed for each candidate processor, and the task is scheduled on the one that minimizes this value.

In the original HEFT implementation~\cite{topcuoglu2002performance}, the communication cost of an edge \((i, k)\) depends solely on the startup latency \(L\) of the processor \(m\) initiating the transfer. This model assumes direct processor-to-processor communication without intermediate memory.  
We extend this model by: (1) incorporating NUMA-aware concepts; (2) explicitly modeling non-uniform communication latencies; and (3) accounting for processor-memory interactions. These extensions align with the modifications introduced in the Min-Min implementation for computing the Earliest Finish Time (EFT), as both algorithms rely on the same EFT formulation (see Figure~\ref{fig:nflow_class_diagram_h}).  
As a result, the updated Equations~\ref{eq:comm_cost_original}, \ref{eq:comm_cost_numa}, and \ref{eq:eft} are also applicable to HEFT.

Finally, our implementation introduces an additional consideration in the calculation of the average communication cost of an edge.  
First, we compute the average latency and bandwidth from the distance matrices, \(L\) and \(B\), respectively, obtained using tools such as the Intel Memory Latency Checker.  
Then, for a given edge, the average communication cost is computed as follows:

\begin{equation}
\overline{c_{i,j}} = \overline{L} + \frac{\text{data}_{i,j}}{\overline{B}}
\label{eq:comm_cost}
\end{equation}

\subsubsection{First-In-First-Out (FIFO)}
A FIFO queue manages ready tasks. When a task releases multiple successors, the newly released tasks are ordered using a level-order traversal of the task graph. Each of these tasks is then assigned a data locality score, defined as the total volume (in bytes) of its input data across all NUMA nodes. This score is used to prioritize the newly released tasks, which are inserted at the end of the queue without altering the order of existing tasks. Level-order traversal is preserved when tasks at the same graph level have identical locality scores. The task at the front of the queue is scheduled on the NUMA node that holds the largest share of its input data. When scores are equal, NUMA node priorities are applied and rotated to ensure balanced task distribution across nodes.

\subsection{Scripts}
\label{sec:scripts}

\subsubsection{validate\_offsets}

\begin{figure}
\centering
\begin{lstlisting}[language=yaml, caption={Sample execution trace passing offsets validation}, label={lst:yaml_sample_offsets}]
...
trace:
  ...
  comm_name_read_offsets:
    Task_2->Task_3: {start: 70, end: 80, payload: 10}
    Task_2->Task_4: {start: 70, end: 90, payload: 20}
    Task_1->Task_2: {start: 30, end: 40, payload: 10}
    Task_1->Task_5: {start: 30, end: 50, payload: 20}

  comm_name_write_offsets:
    Task_2->Task_4: {start: 50, end: 70, payload: 20}
    Task_2->Task_3: {start: 50, end: 60, payload: 10}
    Task_1->Task_5: {start: 10, end: 30, payload: 20}
    Task_1->Task_2: {start: 10, end: 20, payload: 10}

  exec_name_compute_offsets:
    Task_3: {start: 80, end: 90, payload: 10}
    Task_4: {start: 90, end: 100, payload: 10}
    Task_2: {start: 40, end: 50, payload: 10}
    Task_5: {start: 50, end: 60, payload: 10}
    Task_1: {start: 0, end: 10, payload: 10}

  exec_name_total_offsets:
    Task_3: {start: 70, end: 90, payload: 10}
    Task_4: {start: 70, end: 100, payload: 10}
    Task_2: {start: 30, end: 70, payload: 10}
    Task_5: {start: 30, end: 60, payload: 10}
    Task_1: {start: 0, end: 30, payload: 10}
\end{lstlisting}
\end{figure}

This algorithm verifies that the time offsets of read, computation, and write operations for each task in \textit{exec- name\_total\_ offsets} match the overall execution interval defined by the task’s start and end time offsets. For each task, the total duration is computed by considering the durations of read operations (\textit{comm\_name\_read\_offsets}), computation time (\textit{exec\_name\_compute-offsets}), and write operations (\textit{comm\_name\_write\_offsets}), as shown in Equation~\ref{eq:duration}.
The equality condition defined in Equation~\ref{eq:equality} must hold for all tasks in \textit{exec\_name\_total\_offsets}.
If this condition fails for any task, it indicates a violation of data dependencies or an error in the time offset calculations performed by the runtime system, particularly by the mapper (either simulation or bare metal).
Listing~\ref{lst:yaml_sample_offsets} depicts an example of an execution trace that satisfies the validation criteria.
% Note that read and write operations assume that all input and output data items are accessed in parallel, and the total access time is determined by the longest individual access.

\begin{equation}
\label{eq:duration}
    dur(t_i) = \max_{t_k \in pred(t_i)} { (r_{k,i}) } + c_{t_i} + \max_{t_k \in succ(t_i)} { (w_{i,k}) }
\end{equation}

\begin{equation}
\label{eq:equality}
\begin{aligned}
    exec\_name\_total\_offsets.end(t_i) = \\ exec\_name\_total\_offsets.start(t_i) + duration(t_i)
\end{aligned}
\end{equation}

\subsubsection{validate\_output}

\begin{figure}
\centering
\begin{lstlisting}[language=yaml, caption={Expected execution order}, label={lst:yaml_sample_order}]
trace:
  exec_name_total_offsets:
    Task_4:
    Task_3:
    Task_5:
    Task_2:
    Task_1:
\end{lstlisting}
\end{figure}

\begin{figure}
\centering
\begin{lstlisting}[language=yaml, caption={Expected NUMA IDs and Core IDs assigment}, label={lst:yaml_sample_ids}]
trace:
  name_to_thread_locality:
    Task_1: {numa_id: 0, core_id: 0}
    Task_2: {numa_id: 1, core_id: 24}
    Task_3: {numa_id: 1, core_id: 24}
\end{lstlisting}
\end{figure}

\begin{figure}
\centering
\begin{lstlisting}[language=yaml, caption={Expected core availabilities and tasks offsets}, label={lst:yaml_sample_avail}]
runtime:
  core_availability:
    0: {avail_until: 90}
    1: {avail_until: 60}
    2: {avail_until: 70}
    3: {avail_until: 100}

trace:
  exec_name_total_offsets:
    Task_3: {start: 70, end: 90, payload: 10}
    Task_4: {start: 70, end: 100, payload: 10}
    Task_2: {start: 30, end: 70, payload: 10}
    Task_5: {start: 30, end: 60, payload: 10}
    Task_1: {start: 0, end: 30, payload: 10}
\end{lstlisting}
\end{figure}

This algorithm compares two YAML files, an output file (such as the runtime trace output file) and an expected reference, to validate their structural and content correctness. 
It recursively checks whether keys, values, and list elements in the output match those in the expected file. 
Additionally, it can verify that keys appear in a specific order for selected sections, if specified via the \texttt{--check-order} argument. If any mismatch in values, missing keys, list lengths, or key ordering is found, it reports the issue and returns a failure status; otherwise, it confirms that the output is valid. 
% This is useful for verifying the correctness of program outputs against reference specifications.
Listings~\ref{lst:yaml_sample_order}, \ref{lst:yaml_sample_ids}, and \ref{lst:yaml_sample_avail} present examples of the expected output files used for validation. Validations include: the execution order of tasks, where the last task in the list is the first executed following a bottom-up traversal; the assignment of specific cores and NUMA node IDs to individual tasks; and the exact core availabilities along with task start and end time offsets.

\subsection{Cases}

\begin{table*}[ht]
\centering
\caption{Min-Min Algorithm Test Cases}
\label{tab:min_min_cases}
\begin{tabular}{|c|p{4.2cm}|p{4.5cm}|p{5cm}|}
\hline
\textbf{Case ID} & \textbf{Validation Criteria} & \textbf{Expected Outcome} & \textbf{System Setup} \\
\hline
Test 1 & Selects task with min completion time, assigns to earliest available core; updates core availability. & All tasks on 4th core (frequency 8); final core availability = makespan = 70. & Core frequencies = [1,2,4,8]; Task sizes = 80,160,320 FLOPs; Uniform memory access; Independent tasks. \\
\hline
Test 2 & Same as Test 1, but considers NUMA effects. & Tasks 1 and 2 on separate cores; Task3 accesses two memory regions (local, and remote); final makespan = 460. & Two cores at 1 Hz in different memory domains; Task sizes = 160, 360, 80; Non-uniform bandwidths (local=4GB/s, remote=2GB/s). \\
\hline
Test 3 & Compares makespan with uniform vs non-uniform memory access. & Uniform access (4 GB/s) reduces makespan to 440 vs 460 in Test 2. & Same as Test 2, but memory bandwidth is uniform across domains. \\
\hline
\end{tabular}
\end{table*}

\begin{table*}[ht]
\centering
\caption{HEFT Algorithm Test Cases}
\label{tab:heft_cases}
\begin{tabular}{|c|p{4.2cm}|p{4.5cm}|p{5cm}|}
\hline
\textbf{Case ID} & \textbf{Validation Criteria} & \textbf{Expected Outcome} & \textbf{System Setup} \\
\hline
Test 1 & Prioritizes tasks by max upward rank; assigns to earliest finish time; updates core availability. & Execution order: Task3 (core 3), Task2 (core 2), Task1 (core 1); makespan = 40. & Core frequencies = [1,2,4,8]; Task sizes = 80,160,320 FLOPs; Uniform memory access. \\
\hline
Test 2 & Same as Test 1, but includes NUMA effects. & Task2 on core 1, Task1 on core 2; Task3 accesses two memory regions (local, and remote); makespan = 460. & Two 1-Hz cores in different memory domains; Non-uniform bandwidths (local=4GB/s, remote=2GB/s); Task sizes = 160, 320, 80 FLOPs. \\
\hline
Test 3 & Compares makespan with non-uniform vs. uniform access. & Uniform access reduces makespan to 440 (from 460 in Test 2). & Same as Test 2 but with uniform bandwidths (4 GB/s). \\
\hline
\end{tabular}
\end{table*}

\begin{table*}[ht]
\centering
\caption{FIFO Algorithm Test Cases}
\label{tab:fifo_cases}
\begin{tabular}{|c|p{4.2cm}|p{4.5cm}|p{5cm}|}
\hline
\textbf{Case ID} & \textbf{Validation Criteria} & \textbf{Expected Outcome} & \textbf{System Setup} \\
\hline
Test 1 & Verifies FIFO order (level-order traversal) when tasks have equal priorities; updates core availability. & Task order: T1 → T2 → T5 → T3 → T4; makespan = 110. & Single 1-Hz core; 5 tasks, each with 10 FLOPs and 10 bytes communication; Uniform memory access. \\
\hline
Test 2 & Selects the first available core when all cores, priorities (i.e., data locality scores) are equal. & Round-robin assignment across 4 cores; makespan = 70. & Same as Test 1 but with 4 cores at 1 Hz. \\
\hline
Test 3 & Tasks are prioritized based on the amount of resident data in memory, favoring the one with the largest share. & Task order: T1 → T5 → T2 → T4 → T3; makespan = 70. & Identical to Test 2, but with modified edges: Task1 → Task5 [size=20] and Task2 → Task4 [size=20], which break the FIFO task selection order to prioritize by data locality scores. \\
\hline
Test 4 & Selects core with the highest locality for a given task. & Task3 is scheduled on the domain where the majority of data resides. & Two 1-Hz cores in different domains; Task2 writes more data to Task3; Non-uniform bandwidths. \\
\hline
\end{tabular}
\end{table*}

Tables \ref{tab:min_min_cases}, \ref{tab:heft_cases}, and \ref{tab:fifo_cases} describe the test cases used to minimally validate the correct functioning of the runtime system with the scheduling algorithms presented in Section \ref{sec:algorithms} and the previously described validation framework. The validation focuses on two main aspects: (i) correct task prioritization and core assignment, and (ii) makespan calculation. Both aspects are evaluated considering NUMA-aware and non-NUMA-aware versions of the algorithms.
Task prioritization and core assignment are validated in both simulation and bare-metal execution, whereas makespan is validated only through the simulation-based mapper.

\begin{figure}[ht]
\centering
\begin{lstlisting}[language={log}, caption={Test cases directory structure}, label={lst:cases_structure}]
tests
  README.MD
  config
    test_fifo_simulation
      config_1.json
      config_2.json
  system
    test_fifo_simulation
      1_bw.txt
      1_lat.txt
      2_bw.txt
      2_lat.txt
  workflows
    test_fifo_simulation
      config_1.dot
      config_2.dot
  expected
    test_fifo_simulation
      config_1.yaml
      config_2.yaml
  output
    test_fifo_simulation
      config_1.yaml
      config_2.yaml
  log
    test_fifo_simulation
      config_1.log
      config_2.log
\end{lstlisting}
\end{figure}

Test case execution is fully automated via \texttt{make}. The user must provide the nFlows configuration (e.g., \texttt{config\_1.json}), system distance matrices (e.g., \texttt{1\_bw.txt} and \texttt{1\_lat.txt}), the workflow in DOT format, and the expected output in YAML, structured as shown in Listing~\ref{lst:cases_structure}.
Test cases are organized in folders prefixed with \texttt{test\_} (e.g., \texttt{test\_fifo\_simulation}, \texttt{test\_fifo\_bare\_metal}), each containing multiple configurations. Each \texttt{config\_*.json} file defines a single test case. Running \texttt{make test\_fifo\_simulation} executes all test cases in the folder, generating corresponding \texttt{output} and \texttt{log} directories. Results are verified against the expected output, with mismatches detailed in the logs.

As an example, Listings~\ref{lst:fifo_test_4_config} to~\ref{lst:fifo_test_4_output} present the configuration, workflow, distance matrices (bandwidth and latency), expected output (for simulation and bare\_metal mappers), and actual output for Test4 of the FIFO scheduling algorithm. This test (as described in Table\ref{tab:fifo_cases}) verifies that the algorithm correctly selects the core with the highest data locality for a given task, particularly under non-uniform memory access (NUMA) conditions.
The distance matrices are configured such that the communication cost depends only on the bandwidth, which is non-uniform across the NUMA domains. The workflow's compute (FLOPs) and communication payload (bytes) are set to easily produce integer numbers when estimating execution time (compute + communication). This allows for a proper update of core availabilities.
The workflow represents a reduction pattern involving two parallel tasks, Task1 and Task2, each executed in a different NUMA domain where they store a data item locally. The algorithm must then decide on which core to execute Task3, which reduces the data produced by the previous two tasks. Since Task2 writes the largest data item (20 bytes) to the second NUMA node, the algorithm is expected to prioritize a core on that node to maximize data locality.
Finally, the expected output file for this test case is used to validate four key aspects of the scheduling behavior: (i) the execution order, Task1 and Task2 run in parallel, followed by Task3; (ii) task offsets; (iii) selected cores; and (iv) core availabilities. The execution order is verified using the \texttt{trace.exec\_name\_total\_offsets} key, which lists tasks in the order they are scheduled, from bottom (first) to top (last). Offsets confirm that Task3 starts only after Task1 and Task2 complete. Core selection and availability are reported under the \texttt{runtime.core\_availability} key, where the highest availability value corresponds to the workflow’s makespan. The observed makespan matches the manually calculated value when Task3 is correctly assigned to the NUMA domain with the highest data locality. Listing~\ref{lst:fifo_test_4_expected_bare_metal} provides the expected output file for the bare-metal execution of the test case. In this context, the only aspect being validated is the correct assignment of tasks to specific NUMA nodes and cores. As noted at the beginning of this section, makespan validation is only possible when using the simulation mapper.

The cases were executed on the Chameleon Cloud testbed~\cite{keahey2020lessons}, using a single node from an HPC system. This node features two sockets, each with a multi-core Intel® Xeon® Gold 6240R CPU @ 2.40GHz (24 cores, 2 threads per core). Since the nFlows runtime system relies on hwloc to identify core locality, specifically, the NUMA node ID—the configured test cases are expected to execute correctly on systems with similar topologies, particularly those with the same number of cores per NUMA domain. On systems with different NUMA configurations, the core\_avail\_mask attribute in the test definitions may need to be adjusted to ensure the expected scheduling behavior.

% Finally, the expected output file for this test case is used to validate four key aspects of the scheduling behavior: (i) the execution order—Task1 and Task2 run in parallel, followed by Task3; (ii) task offsets; (iii) selected cores; and (iv) core availabilities. The execution order is verified using the \texttt{trace.exec_name_total_offsets} key, which lists tasks in the order they are scheduled, from bottom (first) to top (last). Offsets confirm that Task3 starts only after Task1 and Task2 complete. Core selection and availability are reported under the \texttt{runtime.core_availability} key, where the highest availability value corresponds to the workflow’s makespan.

% Test case execution is fully automated via \texttt{make}. The user must provide: the nFlows configuration file (\texttt{config.json}); system configuration files with bandwidth and latency distance matrices (\texttt{bw.txt} and \texttt{lat.txt}); the workflow in DOT format; and the expected output in YAML, following the structure in Listing~\ref{lst:cases_structure}.
% For instance, running \texttt{make test\_fifo\_simulation} executes all \texttt{config\_*.json} files under the \texttt{config} folder, generating corresponding \texttt{output} and \texttt{log} folders. The command reports whether each test meets the expected result; in case of a mismatch, the logs indicate the source of the error. Multiple configurations (i.e., test cases) can be included in a single \texttt{test\_*} folder.

\begin{figure}
\centering
\begin{lstlisting}[language=json, caption={ FIFO | Test 4 | Config}, label={lst:fifo_test_4_config}]
{
    "dag_file": "./tests/workflows/test_fifo_simulation/config_4.dot",

    "scheduler_type": "fifo",
    "scheduler_params": [
        "fifo_prioritize_by_core_id=yes",
        "fifo_prioritize_by_exec_order=yes"
    ],

    "mapper_type": "simulation",
    "mapper_mem_policy_type": "default",
    "mapper_mem_bind_numa_node_ids": [],

    "core_avail_mask": "0x1000001",
    "flops_per_cycle": 1000000,
    "clock_frequency_type": "static",
    "clock_frequency_hz": 1,

    "distance_matrices": {
        "latency_ns": "./tests/system/test_fifo_simulation/4_lat.txt",
        "bandwidth_gbps": "./tests/system/test_fifo_simulation/4_bw.txt"
    },

    "out_file_name": "./tests/output/test_fifo_simulation/config_4.yaml"
}
\end{lstlisting}
\end{figure}

\begin{figure}
\centering
\begin{lstlisting}[language=dot, caption={FIFO | Test 4 | Workflow}, label={lst:fifo_test_4_workflow}]
strict digraph {
    root    [size=2]; // Ignored in processing.
    end     [size=2]; // Ignored in processing.

    Task_1  [size=10];
    Task_2  [size=10];
    Task_3  [size=10];

    root -> Task_1  [size=2]; // Edge ignored.
    root -> Task_2  [size=2]; // Edge ignored.
    
    Task_1 -> Task_3  [size=10];
    Task_2 -> Task_3  [size=20];

    Task_3 -> end   [size=2]; // Edge ignored.
}
\end{lstlisting}
\end{figure}

\begin{figure}[ht]
\centering
\begin{lstlisting}[language=log, caption={FIFO | Test 4 | Bandwidth matrix}, label={lst:fifo_test_4_bw}]
2
0.005 0.002
0.002 0.005
\end{lstlisting}
\end{figure}

\begin{figure}[ht]
\centering
\begin{lstlisting}[language=log, caption={FIFO | Test 4 | Latency matrix}, label={lst:fifo_test_4_lat}]
2
0 0
0 0
\end{lstlisting}
\end{figure}

\begin{figure}
\centering
\begin{lstlisting}[language=yaml, caption={FIFO | Simulation | Test 4 | Expected output}, label={lst:fifo_test_4_expected}]
runtime:
  core_availability:
    0: {avail_until: 12}
    24: {avail_until: 29}

trace:
  exec_name_total_offsets:
    Task_3: {start: 14, end: 29, payload: 10}
    Task_2: {start: 0, end: 14, payload: 10}
    Task_1: {start: 0, end: 12, payload: 10}
\end{lstlisting}
\end{figure}

\begin{figure}
\centering
\begin{lstlisting}[language=yaml, caption={FIFO | Bare metal | Test 4 | Expected output}, label={lst:fifo_test_4_expected_bare_metal}]
trace:
  name_to_thread_locality:
    Task_1: {numa_id: 0, core_id: 0}
    Task_2: {numa_id: 1, core_id: 24}
    Task_3: {numa_id: 1, core_id: 24}
\end{lstlisting}
\end{figure}

\begin{figure}
\centering
\begin{lstlisting}[language=yaml, caption={FIFO | Test 4 | Output}, label={lst:fifo_test_4_output}]
user:
  flops_per_cycle: 1e+06
  clock_frequency_type: static
  clock_frequency_hz: 1
  distance_lat_ns:
    - [0, 0]
    - [0, 0]
  distance_bw_gbps:
    - [0.005, 0.002]
    - [0.002, 0.005]

workflow:
  execs_count: 3
  reads_count: 2
  writes_count: 2

runtime:
  threads_checksum: 0
  threads_active: 0
  tasks_active_count: 3
  reads_active_count: 2
  writes_active_count: 2
  core_availability:
    0: {avail_until: 12}
    24: {avail_until: 29}

trace:
  name_to_thread_locality:
    Task_3: {numa_id: 1, core_id: 24, voluntary_cs: 0, involuntary_cs: 0, core_migrations: 0}
    Task_2: {numa_id: 1, core_id: 24, voluntary_cs: 0, involuntary_cs: 0, core_migrations: 0}
    Task_1: {numa_id: 0, core_id: 0, voluntary_cs: 0, involuntary_cs: 0, core_migrations: 0}

  numa_mappings_write:
    Task_2->Task_3: {numa_ids: [1]}
    Task_1->Task_3: {numa_ids: [0]}

  numa_mappings_read:
    Task_2->Task_3: {numa_ids: [1]}
    Task_1->Task_3: {numa_ids: [0]}

  comm_name_read_offsets:
    Task_2->Task_3: {start: 14, end: 18, payload: 20}
    Task_1->Task_3: {start: 14, end: 19, payload: 10}

  comm_name_write_offsets:
    Task_2->Task_3: {start: 10, end: 14, payload: 20}
    Task_1->Task_3: {start: 10, end: 12, payload: 10}

  exec_name_compute_offsets:
    Task_3: {start: 19, end: 29, payload: 10}
    Task_2: {start: 0, end: 10, payload: 10}
    Task_1: {start: 0, end: 10, payload: 10}

  exec_name_total_offsets:
    Task_3: {start: 14, end: 29, payload: 10}
    Task_2: {start: 0, end: 14, payload: 10}
    Task_1: {start: 0, end: 12, payload: 10}
\end{lstlisting}
\end{figure}

\section{Limitations}

Although the runtime system is currently in a stable operational phase, some limitations have been identified.
First, at the time of writing, the system has only been tested on low-scale NUMA systems, specifically Chameleon Cloud instances with a maximum of four NUMA nodes. Further testing on systems with a higher number of NUMA nodes is required to evaluate scalability and performance under more complex memory topologies.
Second, the system assumes that scheduling algorithms are dynamic and implement two specific interfaces, \texttt{next()} and \texttt{has\_next()}, to enable incremental task dispatching and core binding. This design constraint may limit compatibility with static scheduling strategies.
Third, the tracing and data collection infrastructure is implemented entirely within the runtime system, without integration of low-overhead, system-level tracing frameworks such as LTTng (Linux Trace Toolkit Next Generation) \cite{LTTngano48:online}. As a result, the overhead introduced by internal instrumentation may become significant when executing large workflows.
Fourth, the current scheduling algorithms iterate over all computing cores and workflow tasks, which may introduce noticeable delays when targeting systems with a large number of cores and workflows with large task counts.
Finally, while the current trace output format, YAML, is practical for small to medium-sized workflows, it becomes increasingly impractical as workflow size grows. The adoption of the Common Trace Format (CTF), a standardized binary format optimized for performance and scalability, is a promising alternative for future integration.

\section{Conclusion and Future Work}

This paper presented nFlows (NUMA-Aware Workflow Execution Runtime System), a tool designed for modeling, executing (via both simulation and on actual hardware), and analyzing scientific workflow scheduling algorithms. A key focus of nFlows is the consideration of Non-uniform Memory Access (NUMA) effects within High-Performance Computing (HPC) systems. We detailed the design and implementation of nFlows, along with its validation and identified limitations.
The development of nFlows opens several avenues for future work, which can be categorized as follows:

\subsubsection*{Extended Support for Memory Policies in Simulation} The current nFlows simulation utilizes the first-touch memory policy. Future iterations will expand this support to include other memory policies, such as bind and next-touch. This will allow for more comprehensive and accurate emulation of diverse memory management strategies.

\subsubsection*{Execution Trace Reproduction} A crucial enhancement consists in enabling the reproduction of execution traces, particularly data access patterns, collected during real executions within the simulation environment. This capability will support the calibration of simulation models and enable detailed analysis of specific access behaviors, thereby assisting performance evaluation and optimization efforts.

\subsubsection*{Advanced Trace Analysis} Further analysis of the generated traces is planned. Since these traces allow for the reconstruction of workflow execution by incorporating bare-metal data access costs, advanced analytical techniques such as Graph Centrality Metrics (from graph theory) will be applied. This analysis is expected to be relevant for identifying potential optimizations and performance bottlenecks within the workflows.

\subsubsection*{Congestion Management Focused Scheduling \cite{gaud2015challenges}} Future work will involve developing scheduling algorithms that prioritize congestion management, rather than exclusively focusing on improving data locality. These algorithms will aim to distribute data access requests from memory-intensive tasks across multiple NUMA nodes to mitigate congestion.

\subsubsection*{In-Memory Workflow Execution Scheduling \cite{kulagina2025memory}} We also plan to develop specialized scheduling algorithms for scientific workflows designed to be executed entirely in memory, leveraging advancements in in-memory computing technologies.

\section{Availability of Dada and Materials}
\textbf{nFlows} is an open-source project. The version employed in this paper, v4.6.1, along with all the test cases described, is available at~\url{https://doi.org/10.5281/zenodo.16749046}. 

%%
%% The acknowledgments section is defined using the "acks" environment
%% (and NOT an unnumbered section). This ensures the proper
%% identification of the section in the article metadata, and the
%% consistent spelling of the heading.
\begin{acks}
Results presented in this paper were obtained using the Chameleon testbed supported by the National Science Foundation.
\end{acks}

%%
%% The next two lines define the bibliography style to be used, and
%% the bibliography file.
\bibliographystyle{ACM-Reference-Format}
\bibliography{sample-base}

\appendix

\end{document}